\documentstyle[12pt]{article}
\begin{document}

\baselineskip=20pt

\title{Moving boundary effects in Quark Gluon Plasma}

\author{J. Seixas$^1$ and J.T.Mendon\c{c}a$^2$
\\ \\$^1$ Centro de F\'{i}sica das Interac\c{c}\~{o}es
Fundamentais, and \\$^2$GoLP/Centro de F\'{\i}sica de Plasmas,
Instituto Superior T\'{e}cnico,\\ 1049-001 Lisboa, Portugal}

\date{}

\maketitle

\begin{abstract}

We analyze the known moving boundary effects in plasma and briefly
review a mechanism for particle acceleration through a plasma
front. We then show how this mechanism may affect particles
crossing the boundary of an expanding quark gluon plasma.

%\vskip 2cm \noindent {\bf Draft paper: not for distribution}

\end{abstract}

\newpage

\section{Introduction}

\bigskip

\bigskip

Heavy ion collisions have been studied all over the years, and
their current interest lies mainly on the search for the
Quark-Gluon Plasma (QGP) state \cite{csernai}. Information on the
eventual transition of the hot nuclear medium to the plasma state
can be obtained from the emission of energetic particles
(electrons and muons) resulting from the heavy ion collision
\cite{abreu}.

Usually it is assumed that particles coming out of the plasma region have their energy
equal to the one they had inside the QGP blob.
However, due to the collective
interactions with the surrounding plasma medium, the final
particle energy can significantly change when they move across the
boundary of the expanding plasma. This energy shift is due, not
only to the difference in the value of the effective potential
created by all the surrounding particles on each side of the
boundary, but also to the velocity of the moving boundary.

This collective effect of the moving background medium has been
explored by several authors in a different context, and
constitutes the basis for plasma acceleration. In this respect,
photon acceleration by relativistic plasma perturbations is a well
defined concept \cite{book}, which has been tested experimentally
with success \cite{dias}. A similar problem, suggested by Bethe
\cite{bethe}, involves the neutrino plasma interactions, where the
electromagnetic force is replaced by the weak force and can
eventually be relevant to dense plasmas in supernovae
\cite{luis,mend}.

From the point of view of the QGP which might have, or has
already, been produced in high energy heavy ion collisions, this
type of collective processes can have a direct consequence, for
instance, on the behaviour of dileptons produced in the plasma
region. For them, QGP is a normal high temperature
electromagnetic plasma and therefore one can approach the process
with techniques previously developed for this kind of medium.

\newpage

\section{Energy states across a plasma boundary}

\bigskip

\bigskip

Let us start with the dispersion relation of a particle moving in
a QGP (in units $c = \hbar = 1$):

\begin{equation}
(E - V)^2 = p^2 + m^2 \label{eq:1} \end{equation} where $p$ is
the particle momentum, $E$ its energy, $m$ its mass and $V$ is the
effective potential created by the interactions with the
surrounding particles of the plasma medium. From here we can
obtain the wave equation  for the particle normalized wave
function:

\begin{equation}
\left( \nabla^2 - m^2 - \frac{\partial^2}{\partial t^2} \right)
\phi = 2 i V \frac{\partial \phi}{\partial t} - V^2 \phi
\label{eq:2} \end{equation}

We shall be interested in a plasma density perturbation moving
with a constant velocity $\vec{v}_s$. For simplicity, we assume
that the perturbation is uniform in the plane perpendicular to
$\vec{v}_s$. We can then choose the $x$ coordinate parallel to
that velocity. Using the d'Alembert transformations to go to the
moving reference frame:

\begin{equation}
\xi = x - v_c t \quad , \quad \eta = x + v_s t \label{eq:3}
\end{equation}
and performing a Fourier transformation in $\eta$, defined as:

\begin{equation}
\phi (\xi, \eta) = \int \phi_q (\xi) e^{i q \eta} \frac{d q}{2
\pi} \label{eq:4} \end{equation} we obtain:

\begin{eqnarray}
\left[ (1 - v_s^2) \left( \frac{\partial^2}{\partial \xi^2} - q^2
\right) + 2 i q (1 + v_s^2) \frac{\partial}{\partial \xi} - m^2
\right] \phi_q (\xi) = \nonumber
\\ = 2 i q v_s V(\xi) \left(i q - \frac{\partial}{\partial \xi}
\right) \phi_q (\xi) + V^2 (\xi) \phi_q (\xi) \label{eq:5}
\end{eqnarray}

If we consider that we are in an equilibrium QGP phase, then we
must accept that the lifetime of the system is larger than the
relaxation times associated with the processes occurring in the
medium. In particular, we must assume that the potential $V (\xi)$
is a slowly varying function. If this is true, we can use the WKB
approximation and set:

\begin{equation}
\phi_q (\xi) = \phi_{q0} \exp \left(i \int^\xi p(\xi') d \xi'
\right) \label{eq:6} \end{equation}

>From this we can get the local dispersion relation:

\begin{eqnarray}
(1 - v_s^2) [p^2 (\xi) + q^2] + 2 (1 + v_s^2) q + m^2 = \nonumber
\\ = 2 v_s V (\xi) [q - p (\xi)] + V^2 (\xi) \label{eq:7}
\end{eqnarray}

We now take the usual definitions:

\begin{equation}
k = - i \frac{\partial}{\partial x} \quad , \quad E = \omega = i
\frac{\partial}{\partial t} \label{eq:8} \end{equation}

We can then establish the relations between the local frequency
(or energy) $\omega (\xi)$, the local momentum $k (\xi)$, the
local variable $p (\xi)$ and q:

\begin{equation}
2 p (\xi) = k (\xi) + \frac{\omega (\xi)}{v_s} \nonumber
\end{equation}

\begin{equation}
2 q = k (\xi) - \frac{\omega (\xi)}{v_s} \label{eq:9}
\end{equation}

This allows us to write the dispersion relation (\ref{eq:7}) in
the form:

\begin{equation}
k^2 (\xi) + m^2 - \omega^2 (\xi) = - 2 \omega (\xi) V (\xi) + V^2
(\xi) \label{eq:10} \end{equation}

And, solving for $\omega(\xi)$, we get:

\begin{equation}
\omega (\xi) = V (\xi) \pm \sqrt{k^2 (\xi) + m^2} \label{eq:11}
\end{equation}

Now, by definition, $q$ is a constant parameter, which for a given
quantum state is invariant inside and outside the plasma region.
Thus, we can quite easily relate the initial values of the energy
and momentum, $\omega_i = \omega (\xi_i)$ and $k_i = k (\xi_i)$,
for a particle created inside the plasma, with the final values
$\omega_f$ and $k_f$, observed outside:

\begin{equation}
2 q v_s = k_i v_s - \omega_i = k_f v_s - \omega_f \label{eq:12}
\end{equation}

The energy shift of the particle will then be:

\begin{equation}
\Delta \omega = \omega_f - \omega_i = (k_f - k_i) v_s
\label{eq:13} \end{equation}

But the initial and final momenta, $k_i$ and $k_f$ can also be
expressed in terms of the energies, according to equation
(\ref{eq:10}). Assuming that the effective potential is equal to
zero outside the plasma, $V_f = 0$, because there are no
interacting particles, we get:

\begin{eqnarray}
k_i &=& \omega_i \sqrt{(1 - V_i/\omega_i)^2 - m^2/\omega_i^2}
\nonumber\\
k_f &=& \omega_f \sqrt{1 - m^2/\omega_f^2} \label{eq:14}
\end{eqnarray}

Replacing this in equation (\ref{eq:12}), we obtain:

\begin{equation}
\omega_f = \omega_i \frac{1 - v_s \sqrt{(1 - V_i/\omega_i)^2 -
m^2/\omega_i^2}}{1 - v_s \sqrt{1 - m^2/\omega_f^2}} \label{eq:15}
\end{equation}

In order to get an explicit result for the final energy, let us
assume that the rest mass of the particle $m$ is negligible, which
is certainly true for the electrons. The results is then:

\begin{equation}
\omega_f \simeq \omega_i \left[ \frac{1 - v_s (1 -
V_i/\omega_i)}{1 - v_s} \right] \label{eq:16} \end{equation}

Which means that the energy shift (\ref{eq:13}) is simply given
by:

\begin{equation}
\Delta \omega \simeq \omega_i \frac{v_s}{1 - v_s} \label{eq:17}
\end{equation}

This means that, when $v_s$ tends to $1$, the energy shift can be
considerable, even if the initial effective potential inside the
plasma is a small correction to the energy of the ejected
particle, $V_i \ll \omega_i$.

Notice that if $v_s < 0$, that is, if we have an imploding surface, leptons traversing it will 
lose energy. For $v_s \sim 1$ leptons can change their energy by as much as half their value inside the 
plasma region.

\newpage

\section{Particle dynamics}

\bigskip

\bigskip

A few remarks should be made concerning the above result. First,
equations (\ref{eq:15} - \ref{eq:17}) are only valid for $V_f = 0$,
which means that the particle has to leave the plasma region before it
spreads away. Second, the exactly resonant case, corresponding to
$v_s = 1$ is not determined by these equations because in this
case the particle will never cross the boundary and, as a result,
its energy remains constant: $\Delta \omega = 0$. Finally, the
above model of a constant velocity $v_s$ can only be relevant if
the changes in the plasma main parameters, such as density and
temperature, are negligible during the interaction time with the
moving boundary.

These, and other, aspects of the problem can be elucidated by using
the classic equations of motion for the emerging particle. They
can be used in the same spirit of the WKB approximation. These
equations can be stated as:

\begin{equation}
\frac{d \vec{r}}{d t} = \frac{\partial \omega}{\partial \vec{k}}
\quad , \quad \frac{d \vec{k}}{d t} = - \frac{\partial
\omega}{\partial \vec{r}} \label{eq:18} \end{equation}

Here, $\vec{r}$, $\vec{k}$ and $\omega$ are the mean values of the
position, momentum and energy of the particle. In a one
dimensional model, we can write:

\begin{equation}
\frac{d x}{d t} = \frac{\partial \omega}{\partial k} = \frac{2
k}{\sqrt{k^2 + m^2}} \nonumber \end{equation}

\begin{equation}
\frac{d k}{d t} = - \frac{\partial \omega}{\partial x} = -
\frac{\partial V}{\partial x} \label{eq:19} \end{equation}

Notice that the energy $\omega$ will change when the particle
moves across the boundary, according to:

\begin{equation}
\frac{d \omega}{d t} = \frac{\partial \omega}{\partial t} =
\frac{\partial V}{\partial t} \label{eq:20} \end{equation}

For a boundary moving without deformation with velocity $v_s$, we
can use $V(x, t) = V(x - v_s t)$, which leads to:

\begin{equation}
\frac{\partial V}{\partial x} = - \frac{1}{v_s} \frac{\partial
V}{\partial t} \label{eq:21} \end{equation}

Comparing with equations (\ref{eq:19}, \ref{eq:20}), we conclude
again that $(\omega - k v_s)$ is a constant of motion, as
considered before.

Let us consider the case where $V(x, t) \neq V (x - v_s t)$, and
this constant of motion is not valid. As a first example we can use:

\begin{equation}
V (x, t) = V_0 (t) \frac{1}{2} \left\{ 1 + \tanh \left[ k_s (x -
x_0 (t) \right] \right\} \label{eq:22} 
\end{equation}
where $k_s$
defines the slope of the boundary, and $x_0 (t) = x_0 + v_s t$ is
its mid position. Here, we can also define the amplitude of the
potential $V_0 (t)$ as nearly proportional to the mean plasma density.
Actually, there is also a small dependence on the plasma
temperature, associated with Debye screening. If the plasma is
initially inside a sphere of radius $r_0$, and it is expanding along
the axis $x$, we can then approximately write:

\begin{equation}
V_0 (t) = V_0 \frac{1}{1 + (v_s / r_0) t} \label{eq:23}
\end{equation}

The energy shift of the particle moving across the boundary along
the $x$ axis can be determined by:

\begin{equation}
\omega (x, t) = V (x, t) + \sqrt{k^2 (t) + m^2} \label{eq:24}
\end{equation}
where the particle postion $x (t)$ and momentum $k (t)$ are
determined by the above equations of motion.

A more realistic example is provided by the expanding fireball model \cite{Rapp} for which we assume
a cylindrical expanding plasma region such that the volume evolves like
\begin{equation}
V_0 (t)=2\pi\left\{x_0 +v_s t +\frac{1}{2}a_x t^2\right\}(\tau_0+\frac{1}{2}a_{\perp}t^2)^2
\end{equation}
where the parameters are fixed using flow data. Typical values are $\tau_0\sim 4.6$ fm an a total freezout time 
of $t_f \sim 10-12$ fm in the longitudinal $x$ direction and $3-4$ fm along the perpendicular direction, 
which implies a final velocity of the front along the $x$ axis, $v_x \sim 0.75 c$ and $v_{\perp}\sim 0.55c$. In this case, for two particles with the same energy inside the plasma $\omega_i$, but crossing different plasma boundaries we have the relation
\begin{equation}
\frac{\omega_{f_\perp}}{\omega_{f_{||}}}\sim\frac{1-v_{s{||}}}{1-v_{s{\perp}}}.\frac{1-v_{s{\perp}}(1-V_i / \omega_i)}{1-v_{s{||}}(1-V_i / \omega_i)}
\end{equation}

In order to illustrate the relevance of the collective plasma
effects and to point out the main aspects of the problem of
determining the initial energy of the observed particles, we
present numerical examples resulting from the integration of the
equations of motion. In figure 1, we show the energy shift for two
different values of the front velocity $v_s$, and a constant plasma
potential. We notice that we can obtain a non-negligible energy
shift $\Delta \omega \gg V$, which increases when $v_s$ approaches
one. This illustrates the importance of the collective effects
associated with a rapidly moving potential on the particle dynamics. In
figure 2, we show that such energy shifts can be reduced if the
maximum potential amplitude decreases during the interaction time
with the boundary. Even so, there is still a large range of
parameters for which the energy shifts can be significant.

\newpage

\section{Conclusions}

\bigskip

\bigskip

We have shown that the energy of a particle, for instance, an
electron or a muon, measured during high energy ion collisions and
emerging from the dense nuclear matter gas or quark gluon plasma
can be significantly different than its initial energy. This
results from the influence of the effective potential acting on
that particle as a consequence of its interaction with the
background plasma particles. Even if this effective potential
remains negligible with respect to the initial particle energy,
its influence can be important due to the fact that it is changing
on very fast space and time scales. In particular, the potential
time variation create a force which attains its maximum value at the expanding
plasma boundary. Even if this force is weak, it can act for a long
time if the particle moves along with the boundary. The results
are valid for electromagnetic and for strong interactions with the
background particles. They suggets that the mean field acceleration
processes already known in other other areas of plasma physics can
also be relevant to high energy physics.

This work shows that the resulting energy shifts are very
sensitive to the space and time variations of the expanding
plasma, and in particular to the velocity of its boundary.

In this paper we have used the WKB approximation, which is not
valid if the background plasma potential changes on scales
comparable to the wavelength of the emerging particles. However,
it can easily be realized that a full quantum model will lead to the
same energy shifts. Typical quantum processes not described by the
present model, such as reflection of particles from the boundary
and coupling between different helicity states will be discussed
in a future work.

\bigskip

\bigskip

\textbf{FIGURE CAPTIONS}

\bigskip

Figure 1 - Particle energy as a function of time, for $m^2 = 0.1$,
for a tangent hyperbolic front with $k_s = 1$, and a constant
potential $V_i = 1$. The front velocities are: (a) $v_s = 0.98$,
and (b) $v_s = 0.99$.

\bigskip

Figure 2 - The same as in Figure 1, but for a time dependent
potential, as defined by equation (\ref{eq:23}), with $v_s/r_0 =
0.1$.


\newpage
\begin{thebibliography}{99}

\bibitem{csernai}
L.P. Csernai, {\sl Introduction to Relativistic Heavy ion
collisions}, John Wiley and Sons, New York (1994).

\bibitem{abreu}
M. Abreu et al, {\sl Phys. Lett. B}, {\bf 410}, 337 (1997).

\bibitem{book}
J.T. Mendon\c{c}a, {\sl Theory of Photon Acceleration}, Institute
of Physics Publishing, Bristol, (2001).

\bibitem{dias}
J.M. Dias et al., {\sl Phys. Rev. Lett.}, {\bf 78}, 4773 (1997).

\bibitem{bethe}
H. Bethe, {\sl Phys. Rev. Lett.}, {\bf 56}, 1305 (1986).

\bibitem{luis}
L.O. Silva et al., {\sl Phys. Rev. Lett.}, {\bf 83}, 2703 (1999).

\bibitem{mend}
J.T. Mendon\c{c}a et al., {\sl J. Plasma Phys.}, {\bf 64}, 97
(2000).

\bibitem{Rapp}
R. Rapp, G. Chanfray and J. Wambach, Nucl. Phys. A617 (1997) 472.

\end{thebibliography}
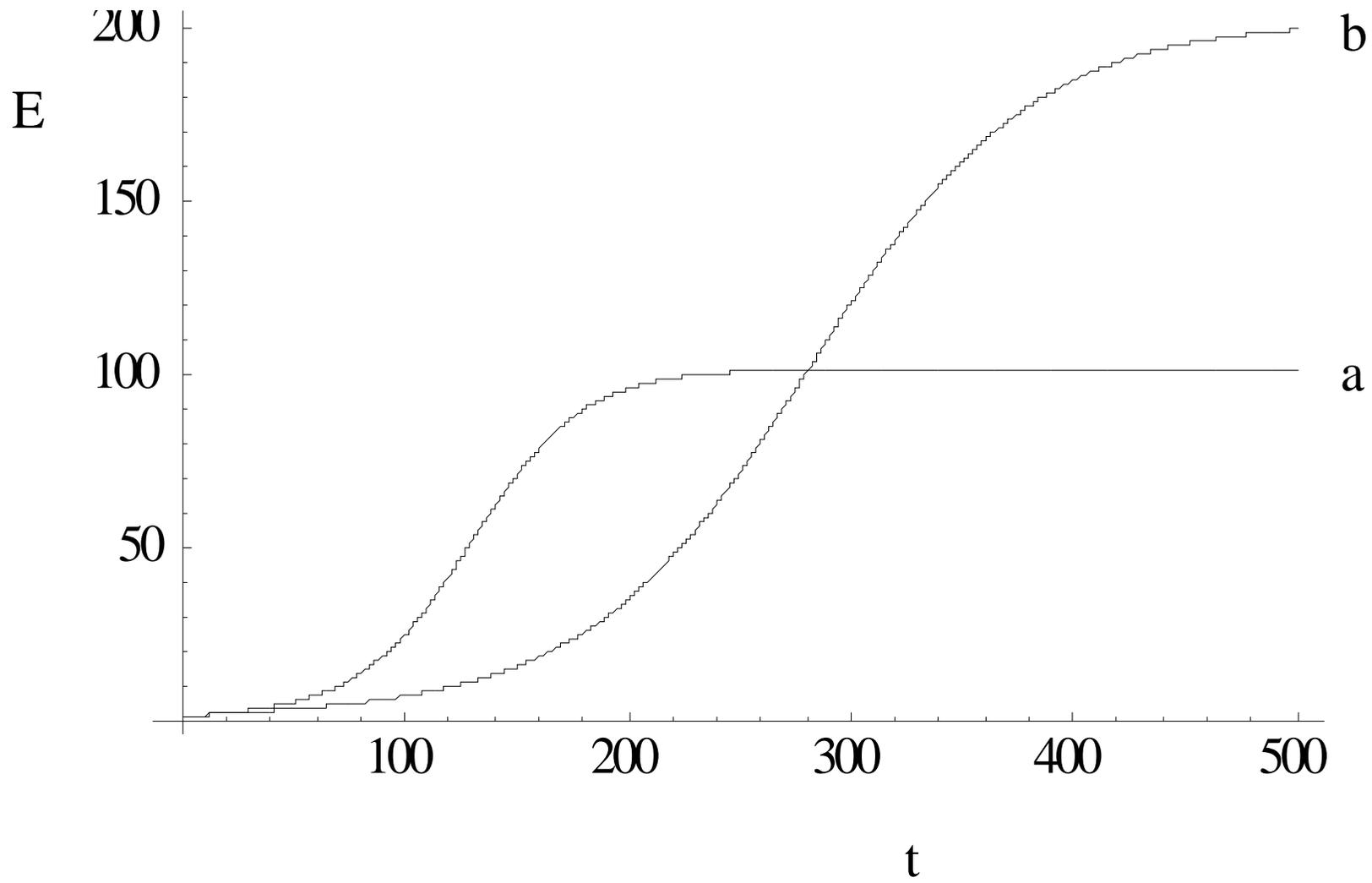
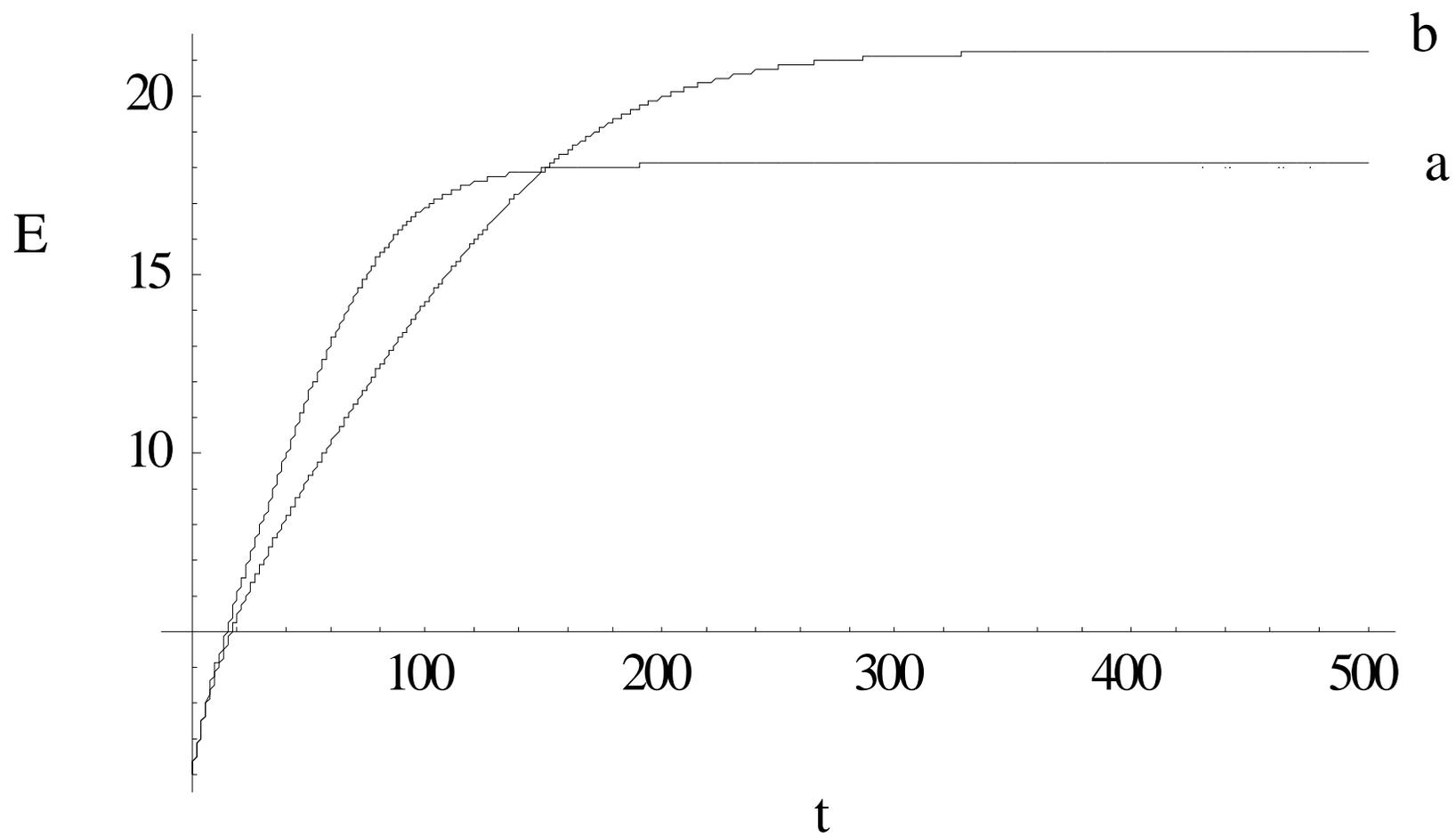
\end{document}